\documentclass[]{aipproc}

\layoutstyle{6x9}

\usepackage{graphicx}	
\usepackage{epstopdf}

\begin{document}

\title{Core-crust transition pressure for relativistic slowly rotating neutron
  stars}


\author{L. M. Gonz\'alez-Romero}{
address = {Depto. F\'{\i}sica Te\'orica II, Facultad de  Ciencias F\' \i sicas,
Universidad Complutense de Madrid, 28040-Madrid.},
email={mgromero@fis.ucm.es}}

\author{J. L. Bl\'azquez-Salcedo}{
address = {Depto. F\'{\i}sica Te\'orica II, Facultad de  Ciencias F\' \i sicas,
Universidad Complutense de Madrid, 28040-Madrid.},
email={joseluis.blazquez@fis.ucm.es}}

\begin{abstract}
We study the influence of core-\textit{crust} transition pressure changes on the
general dynamical properties of neutron star configurations. First we study
the matching conditions in core-\textit{crust} transition pressure region,
where phase transitions in the equation of state causes energy density
jumps. Then using a surface 
\textit{crust} approximation, we can construct configurations where the
matter is described by the equation of state of the core of the star
and the core-\textit{crust} transition pressure. We will consider neutron
stars in the slow rotation limit, considering perturbation theory up to second
order in the angular velocity so that the deformation of the star is also
taken into account. The junction determines the 
parameters of the star such as total mass, angular and
quadrupolar momentum. 
\end{abstract}

\maketitle

\section{Introduction}
The interior of the neutron stars is composed of matter in a highly compact
state. Most of the theories that describe the matter composition of these stars
predicts a layer structure, resulting from the different matter species that
emerge as a result of the extreme densities reached
\cite{glendenning2000compact}. Neutron stars are 
composed by the core and the \textit{crust}, whose properties are very
different. In the core of the star (around 10 Km of radius), matter is found
at densities well beyond nuclear density. The equation of state for matter in
this state is not well 
known, and different hypothesis exist, essentially differing on which particle
populations could be found and which ones dominates over the others. The outer
regions of the core is composed by neutrons, protons, electrons and possibly
muons while in the interior regions more exotic matter states are thought to
be found (pions, hyperons, quark matter...). The
\textit{crust} of neutron stars is reached at
nuclear densities, and a mix of 
neutrons, free electrons, and neutron-rich atomic nuclei is thought to give a
metallic structure to this 
region \cite{haensel2007neutron}. 

Along the core of the star and specially in the core-\textit{crust} interface,
first order phase transitions are expected to be found in 
realistic equations of state. These phase transitions result in small discontinuities in
the energy density of the star matter
\cite{haensel2007neutron,heiselberg-2000-328}. The appropriate way of
treating these discontinuities is 
by analyzing the junction between every region in the interior of the star,
i.e., 
the matching conditions in the core-\textit{crust} transition region. This
junction must satisfy some basic conditions so that the resulting metric is
continuous and the Einstein equation is satisfied. We use the intrinsic
formulation of these conditions that can be written in terms of the first and
second fundamental forms
\cite{darmois1927équations,lichnerowicz1955théories,misner1973gravitation}. We
will make an  
interpretation of this matching in terms of physical quantities like the
transition pressure. 
\section{Matching the transition pressure
  region}
In nature neutron stars are found rotating in the form of
pulsars. Although the angular velocity of these objects is very high (with
periods from seconds to milliseconds), when compared with the gravitational
energy of the compact star, the rotational energy of the star is in most cases
smaller. So that for modeling a realistic neutron star we can use the slow rotation
Hartle-Thorne perturbative theory for rigid rotating axi-symmetric stars. We
will consider the slow rigid 
rotation approximation up to second order in the angular velocity, so that the
deformation of the star due to its rotation is also considered. We will follow
the notation of the original papers by Hartle and Thorne
\cite{1967ApJ...150.1005H,1968ApJ...153..807H}, with the perturbed 
metric given by $ds^{2}=-e^{2\psi}[1+2h]dt^{2}+e^{2\lambda
 }[1+2m]dr^{2}+r^{2}[1+2k]\left[d\theta ^{2}-\sin
  ^{2}\theta(d\phi -\omega dt)^{2}\right]$.

We will consider the case in which a interior region with perfect fluid with a
given equation of state is being matched to an exterior perfect fluid with
another equation of state. The matching will be done on a surface of constant
pressure (up to second order in the angular velocity). The inner face of the
surface will have in general some values $(p_{-},\rho_{-})$, and the
exterior region $(p_{+},\rho_{+})$ \cite{PhysRevD.67.064011}. 

We must grant that the general
solution satisfy the continuity of the first fundamental form in order to have
a well defined metric on the surface. The Einstein equations tell us that the
discontinuity resulting from the jump of the 
second fundamental form can be written as a surface stress-energy
tensor in the core-\textit{crust} interface. So introducing a surface energy
density on the interface surface causes a discontinuity of the pressure on
this surface. We will
write this superficial stress-energy tensor as a perfect fluid tensor with
surface density $\rho _{c}(\theta )=\varepsilon +\delta \varepsilon (\theta
)=\varepsilon +\delta \varepsilon _{0}+\delta \varepsilon _{2}P_{2}(\theta )$ 
where $\varepsilon$ is the zero order surface density and $\delta \varepsilon
_{0}, \delta \varepsilon _{2}$ are the second order spherical and quadrupolar
surface densities. They are all constant on the junction interface.

Some of the matching conditions for the functions of the metric are the
following: 
\begin{eqnarray}
\Delta [e^{-\lambda(a)}] = -4\pi a \varepsilon  
\label{59}\\
\Delta \left[\frac{M(a)+4\pi
  a^{3}p(a)}{\sqrt{1-\frac{2M(a)}{a}}}\right] = 4\pi a^2 \varepsilon 
\label{60}\\
\Delta [\omega (a)]=0
\label{61}\\
\Delta \left[e^{-\lambda(a)}\partial _{R}\overline{\omega}(a)\right]
= \frac{16\pi\varepsilon}{a}(\Omega_{c}-\omega(a))
\label{62}
\end{eqnarray}
 Where $\Delta \lbrack f(a)]=f_{+}-f_{-}$. These
 equations provides the matching for the zero and first order 
functions. Essentially the introduction of a discontinuity in density and
pressure inside of the star implies the apparition of a surface energy density
that also modifies the mass. The inertial dragging must be continuous across
the surface, while the derivative of the inertial dragging presents a
discontinuity given by the angular velocity of the surface density
(which modifies the total angular momentum of the star). If the angular
velocity of the two regions is the same, then the surface density must rotate
with the same angular velocity.

Studying the matching up two second order reveals matching conditions for the
remaining metric functions, providing the jump in the mass perturbation,
quadrupolar moment and second order terms of the surface energy density. These
equations will be presented elsewhere.
\section{Surface \textit{crust} approximation}
\begin{figure}[h]
\resizebox{0.5\textwidth}{!}
{\includegraphics{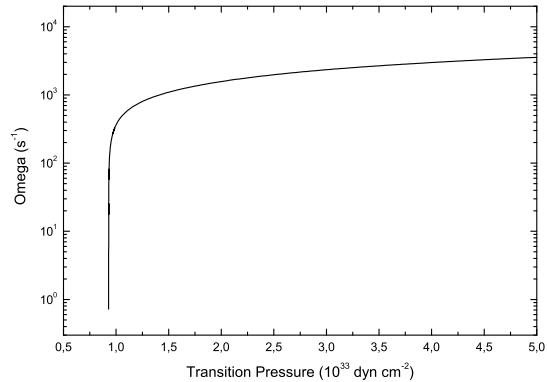}}
\label{fig:1}
\caption{Angular velocity vs transition pressure for central density
  $1.279 \cdot 10^{15} g cm^{-3}$  and $1.44M_{\odot}$}
\label{fig:1}
\end{figure}
In general we will need to integrate, inside the star, the metric functions in
different regions 
separated by the different phase transitions. If we just consider the phase
transition at the core-\textit{crust} interface, we will need to integrate in
two regions, 
inside the core of the 
star and along the \textit{crust}. We will need to impose the previous boundary
conditions that 
grant the appropriate continuity of the metric across the regions. In order to
simplify the 
situation so that only one region needs to be integrated, we can construct a
simplified model of 
\textit{crust} using the obtained matching conditions: 

It can be proved that in the \textit{crust} the
density falls down to zero very quickly, and every contribution from the
\textit{crust} to integral properties of the star  like the mass or the 
angular momentum can be neglected. The
outer \textit{crust} extends only over a region of 
some hundreds of meters with the density rapidly falling to zero. So
approximating the  
\textit{crust} of the star to a surface density enveloping the core of the
neutron star  is a well justified model of \textit{crust} where the obtained
radius can be 
interpreted as an effective radius of integration for the star. It can be seen
that this approximation can be achieved in the following way.

We can integrate the core of the star with an appropriate equation of state,
and match directly to the exterior vacuum solution. Then the 
exterior pressure and density becomes zero in the 
expressions for the matching, and the outer mass function becomes the
mass outside the star. The 
interior pressure can be taken at the inner \textit{crust}. The surface
density obtained from the matching conditions can be interpreted as the
approximated \textit{crust}: the  \textit{crust} becomes a surface
energy density enveloping the core of the star.

This
simplified \textit{crust} model can tell us a lot of information about the
dependence of the properties of the star with the core-\textit{crust}
transition pressure. In Figure \ref{fig:1} we present the relation of the
angular velocity of the neutron star with the transition pressure. These
results are for configurations with total second order mass $1.44M_{\odot}$
and central density $1.279 \cdot 10^{15} g\cdot cm^{-3}$. The equation of state
used for these results is the Glendenning 240 for high densities together with
the BPS equation of state for lower densities \cite{glendenning2000compact}. 

In the figure it can be seen two different regions, one where the angular
velocity is quite sensitive to core-\textit{crust}
transition pressure variations, and another where it is less important. Around
core-\textit{crust} transition pressure other properties of the star like the
angular momentum and the quadrupolar momentum are also heavily modified with
small perturbations of the core-\textit{crust} transition pressure
variation. The specific value for which this variations can affect the global
properties of the star is different for other central densities and equations
of state. In this case, small variations of the order of $10^{23} dyn\cdot cm^{-2}$
causes relative variations order $10^{-6}$ in the period, angular momentum and
eccentricity. 

The configurations have been generated integrating the differential equations
for using a Fortran code based in the Colsys routine
\cite{Ascher:1981:CSB:355945.355950}. The equation of state 
has been interpolated numerically from the tables. The 
interpolation is based on a natural spline that makes the adiabatic index
continuous (necessary to obtain a good precision in the second order
perturbation functions).

Colsys is the appropriated
way of resolving the problem, because Colsys allows to resolve differential
equations with quite general boundary conditions. We impose the usual
regularity conditions together with the boundary conditions at the border of
the star that grants the solution outside the star to be asymptotically flat
with the matching previously explained. Also Colsys allows to
introduce fixed points of the mess where certain junction conditions could be
imposed. This feature will be used in future works to integrate multiple
regions with different equations of state.

The angular velocity of the star is quite dependent of the transition
pressure, specially in the region nearby the core-\textit{crust}
transition. This means that small changes in the core-\textit{crust}
transition pressure could give rise to important changes in the dynamical
properties of the star like angular momentum and quadrupolar
moment. Energy depositions on the outer layers of the star could affect the
phase transition, causing variations in the angular velocity of the star. These
processes could explain pulsar \textit{glitches}.
\bibliography{proc_ERE2011}
\bibliographystyle{aipproc}
\end{document}